\documentclass{ws-mpla}
\usepackage{url}              
\usepackage{amsmath}

\begin{document}

\markboth{Alberto Accardi}
{Large-\emph{x} partons from JLab to the LHC}

\catchline{}{}{}{}{}

\title{LARGE-\emph{x} CONNECTIONS \\ OF NUCLEAR AND HIGH-ENERGY PHYSICS}

\author{\footnotesize Alberto Accardi}

\address{
        Hampton U., Hampton, VA 23668, USA \\ 
        Jefferson Lab, Newport News, VA 23606, USA}

\maketitle


\begin{abstract}
I discuss how global QCD fits of parton distribution functions can make the somewhat separated fields of high-energy particle physics and lower energy hadronic and nuclear physics interact to the benefit of both. 
I review specific examples of this interplay from recent works of the CTEQ-Jefferson Lab collaboration, including hadron structure at large parton momentum and gauge boson production at colliders. 
I devote particular attention to quantifying theoretical uncertainties arising in the treatment of large partonic momentum contributions to deep inelastic scattering observables, and to discussing the experimental progress needed to reduce these.

\end{abstract}

\section{Overview}
\label{sec:overview}

\begin{quote}
{\em
``The coherence provided by QCD means that insights [into hadron 
structure] may arise from unexpected quarters. It is more than ever
advisable to take a broad view that integrates across hadronic physics,
and to connect with the rest of subatomic physics.'' 
} C. Quigg, 2011 \cite{Quigg:2011sb}
\end{quote}

\noindent
The across-the-board connection I will focus on here is between the fields of high-energy particle physics and of lower energy hadronic and nuclear physics. This connection is made possible by the universality of the parton distribution functions (PDFs) of the proton, which allows one to calculate a variety of processes from a common set of quark and gluon momentum distributions. These processes range from high-energy physics interactions such as in $\bar p+p$ collisions at the Tevatron or $p+p$ collisions at the Large Hadron Collider (LHC) to lower energy electron or hadron collisions on proton as well as nuclear targets. The former are designed to study  QCD and electroweak interactions at the parton level, as well as to search for the Higgs boson and for physics beyond the standard model. The latter are designed to explore how QCD builds a hadron out of 3 valence quarks, how its quantum numbers are built up from these and from the ``sea'' of quark-antiquark pairs and gluons, how its properties change in a nuclear medium, and how a nucleus as a bound state of protons and neutrons emerges from the underlying microscopic quark and gluon degrees of freedom. 

One powerful set of tools that exploit the PDF universality are the so called ``global PDF fits\cite{Forte:2013wc,Jimenez-Delgado:2013sma,Rojo:2013fta}''. Their original \emph{ra\^ison d'\^etre} is to utilize experimental data from a number of processes, combined with perturbative QCD calculations of the relevant partonic cross sections, in order to extract the non perturbatively calculable parton distributions. These in turn can be used to calculate processes not included in the fits, for example the expected rates of Higgs boson production in several channels, or various standard or beyond-the-standard model cross sections, such as production of $W'$ and $Z'$ gauge bosons, Kaluza-Klein resonances, gluinos, and so on. In this sense, increasing the number of data points to be fitted by including more processes, and improving the theoretical calculations to include a larger portion of the available kinematics, is extremely useful to reduce the uncertainties in the extracted PDFs, thus yielding more accurate theoretical predictions. This is one way in which lower energy hadronic and nuclear data, which typically access lower momentum scales and larger parton momentum fractions $x$ inside a proton than at colliders, can improve the study of high-energy processes.

However, as I will argue here, this phenomenologically very important connection is not necessarily the most far reaching. Indeed, a novel aspect of large-$x$ global fits is their ability to connect elements of high-energy physics with hadronic and nuclear physics at medium energy. For example, data on $W$ and $Z$ forward rapidity boson production at the Tevatron and LHC, which can reach large values of $x$, can constrain the extrapolation to $x=1$ of the down-quark to up-quark ratio in the proton. With enough statistical precision this can indicate which nonperturbative proton structure model best captures the effects of confinement on hadron structure. Here we have collider physics providing insights on hadronic physics.

Likewise, but less obviously, global PDF fits can be used as a tool to study the structure of the nucleus and the differences between bound and free protons and neutrons, for which current models display large theoretical uncertainties. This is possible, {\it e.g.}, by exploiting the interplay of Deep Inelastic Scattering (DIS) data \emph{on deuteron targets}, which allow one to extract a nuclear-model dependent $d$ quark distribution at large $x$ \cite{Accardi:2009br,Accardi:2011fa,Owens:2012bv}, and weak interaction processes \emph{on proton targets}, such as the mentioned $W$ and $Z$ production at hadronic colliders, which also depend on the $d$ quark but are naturally free from nuclear effects. Since an unrealistic nuclear correction would pull the $d$ quark extracted from a global fit away from that required by the proton target data, constraints on the viable nuclear models can then be obtained by studying tensions in these two data sets under global fits. 
Global fits can thus relate high-energy experiments at the Tevatron
\cite{Grannis-Tevatron,Denisov-Tevatron} or LHC\cite{Dissertori:2012cra} with nuclear physics experiments at lower energy facilities such as Jefferson Lab\cite{JLab10,Dudek:2012vr}. Otherwise stated, one can now use proton targets to study QCD in nuclei. 

It is the goal of this review to substantiate these claims.

\section{The CTEQ-Jefferson Lab global PDF fits}

\begin{figure}
\vskip-.3cm
\parbox{0.6\linewidth}{\center
\includegraphics[width=0.95\linewidth,trim= 0 120 0 25,clip=true]
                {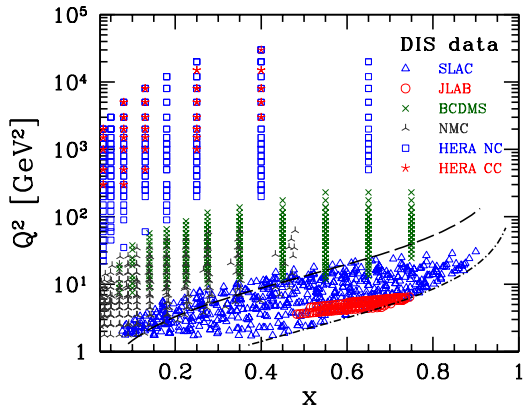}
}
\parbox{0.35\linewidth}{
\caption{Kinematic coverage in Bjorken $x_B$ and $Q^2$ of the DIS data used in the CJ12 fits \protect\cite{Owens:2012bv}. The $W^2 \gtrsim 14$~GeV$^2$ cut typically used in global PDF fits and the $W^2>3$ GeV$^2$ cut used in the CJ12 analysis are indicated by the dashed and dot-dashed lines, respectively.
\protect\label{fig:DISdata} }
}
\vskip-.5cm
\end{figure}

\noindent
Deep-inelastic lepton-hadron scattering experiments have provided
a wealth of data over the past few decades that have yielded
considerable information on PDFs over a wide range of $x$ and $Q^2$.
However, most global fits\cite{Martin:2009iq,JimenezDelgado:2008hf,JimenezDelgado:2009tv,Lai:2010vv,Aaron:2009aa,Ball:2012cx} have focused on the extraction of
leading twist PDFs, utilizing cuts on $Q^2$ and the hadronic final
state mass $W^2 = M^2 + Q^2 (1-x)/x$, where $M$ is the nucleon mass,
of $Q^2 \gtrsim 4$~GeV$^2$ and $W^2 \gtrsim 14$~GeV$^2$.  The aim
of such cuts is to eliminate regions of kinematics where effects
that do not scale with $log(Q^2)$ may be important, which unfortunately
has the effect of excluding a considerable amount of high-precision
data that have been collected at intermediate and large values of $x$, see Figure~\ref{fig:DISdata}.

On the other hand, there are many reasons why accurate information
on PDFs at high $x$ is important.  For example, it is necessary to
have control over uncertainties on QCD backgrounds in searches for
new physics in collider experiments with final states at large $p_T$, large invariant mass, or large rapidity\cite{Rojo:2013fta}.
Also, the behavior of PDF ratios, such as the $d/u$ ratio, as $x \to 1$
can provide insight into the dynamics of quarks and gluons in the
nonperturbative region \cite{Holt:2010vj}.  In addition, the uncertainty in the extraction of the spin-dependent gluon PDF at small $x$ in forward particle
production in polarized $pp$ collisions is limited by the uncertainties
on the quark PDFs at large $x$.

The CTEQ-Jefferson Lab (CJ) collaboration\cite{CJweb} was formed with the initial aim to include large-$x$ and ${\cal O}(1/Q^2)$ theoretical corrections into perturbative calculations, in order to maximize the use of available experimental data (in particular from fixed target DIS experiments) and produce accurate fits with a quantitative evaluation of the associated theoretical errors. Earlier efforts in this direction include the work of Alekhin {\it et al.}\cite{Alekhin:2000ch,Alekhin:2012ig} and Martin {\it et al.}\cite{Martin:2003sk}

\begin{figure}
\vskip-.3cm
\center
\includegraphics[width=0.7\linewidth,trim= 0 230 0 0,clip=true]
                {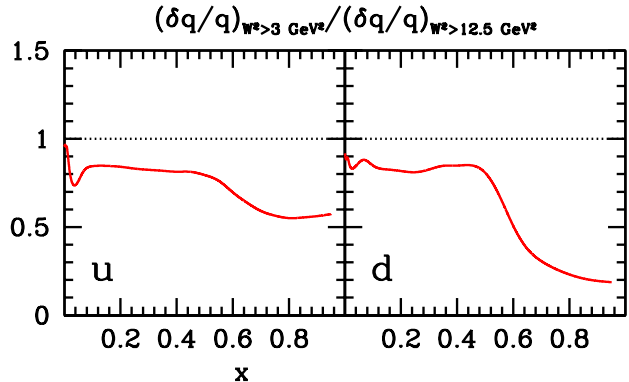}
\caption{Ratio of relative errors of the CJ12mid fit with $W^2>3$ GeV$^2$ and a a similar fit using $W^2>12.5$ GeV$^2$.
\label{fig:improvedstat}
}
\end{figure}   

The CJ global PDF fits utilize the world's data on charged lepton
DIS on proton and deuteron targets, lepton pair production with a
proton beam on proton and deuteron targets, $W$ asymmetries and $Z$ rapidity distribution as well as jet production data in $\overline p p$ collisions.
The theoretical treatment of inclusive DIS includes subleading ${\cal O}(1/Q^2)$ effects, such as target mass correction and (a fit of) higher twist terms, and nuclear corrections for the deuterium target data.
The resulting fits\cite{Accardi:2009br,Accardi:2011fa,Owens:2012bv} can thus incorporate data down to $W \approx 1.7$~GeV, and have culminated in the release of the ``CJ12'' PDF sets\cite{CJweb,CJ-LHAPDF}, valid in the
range $10^{-5} \lesssim x \lesssim 0.9$.
The fits were performed at next-to-leading order in the zero
mass variable flavor number scheme, with $\alpha_s(M_Z)$ fixed to the
world average value.
(Full heavy quark treatments, fits of the strong coupling constant,
and inclusion of the available LHC data will be considered in a
subsequent analysis.)

The CJ PDFs have been shown\cite{Accardi:2009br} to be stable with the weaker cuts on $W$ and $Q^2$, and the increased DIS data sample (of about 1000 additional points) has led to significantly reduced uncertainties of up to 80\% on the $d$ quark PDF at large $x$, where precise data are otherwise scarce (see figure \ref{fig:improvedstat}). Since $d$ quark flavor separation at large $x$ is currently almost entirely dependent on DIS on deuterium targets, corrections for nuclear Fermi motion and binding effects are included by convoluting the nucleon structure functions with a smearing function computed from the deuteron wave function \cite{Accardi:2011fa,Kulagin:2004ie}.  As the $u$ quark is well constrained by data on proton targets, the $d$ quark becomes directly sensitive to the nuclear corrections. The effect is a large suppression at high $x$, and a mild but non-negligible increase at intermediate $x$, still inside the ``safe'' region defined by the larger $W$ cut discussed above\cite{Accardi:2009br}.  These findings have subsequently been confirmed by Ball {\it et al.}\cite{Ball:2013gsa}.

\begin{figure}
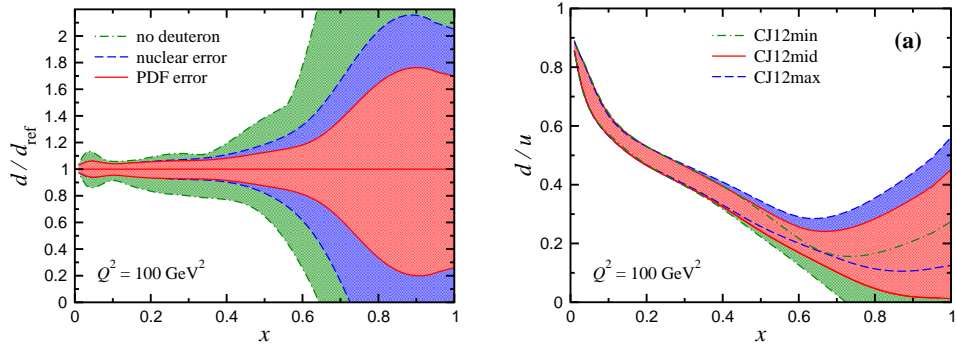

\center
  \includegraphics[width=0.47\linewidth,trim= 0 0 0 0,clip=true]
                  {relerrNODd.eps}
  \hspace{.5cm}
  \includegraphics[width=0.462\linewidth,trim= 0 0 0 0,clip=true]
                  {relerrCJdu.eps}
  \caption{
  {\it Left:} PDF uncertainties for the CJ12mid $d$ quark
	compared with the total uncertainty including nuclear
	corrections and with a fit excluding all deuterium
	data relative to the CJ12mid set. 
  {\it Right:} $d/u$ ratio for the CJ12min, CJ12mid and CJ12max PDFs
	\protect\cite{Owens:2012bv}. Note that these ratios are to 
        a good approximation independent of the $Q^2$ scale.
\label{fig:CJ12d}
}
\end{figure}

The uncertainties on the CJ12 $d$ quark PDF from theoretical modeling of
nuclear corrections (which I refer to as ``nuclear uncertainties'')
have been quantified\cite{Accardi:2011fa,Owens:2012bv}.
In particular, the size of the nuclear corrections range from mild, corresponding to the hardest of the deuteron
wave functions (WJC-1) coupled to a 0.3\% nucleon off-shell
correction, to strong, corresponding to the softest wave function
(CD-Bonn) and a large, 2.1\% nucleon off-shell correction; the
central value corresponds to the AV18 deuteron wave function with
a 1.2\% off-shell correction\cite{Owens:2012bv}.  The resulting PDFs
are labeled ``CJ12min'', ``CJ12max'', and ``CJ12mid'', respectively.
The uncertainties on the CJ12 $d$ quark distribution are illustrated
in the left panel of Figure~\ref{fig:CJ12d}, with the central (red) band indicating the PDF error calculated with the Hessian method and a tolerance
factor $T=10$.  The central (blue) band represents the theoretical
nuclear uncertainty, obtained as an envelope of the CJ12min and
CJ12max fits, and is of the same order of magnitude as the PDF error.
Finally, the outer (green) band is the PDF error in a fit that excludes 
all deuterium data, and exceeds the combined PDF and nuclear uncertainties.
This demonstrates the usefulness of the deuterium data, even in
the presence of the nuclear uncertainties that its use introduces.

\begin{figure}
\center
\vskip-.3cm
\parbox{0.9\linewidth}{
\includegraphics[width=\linewidth,bb=18 330 592 718,clip=true]
		{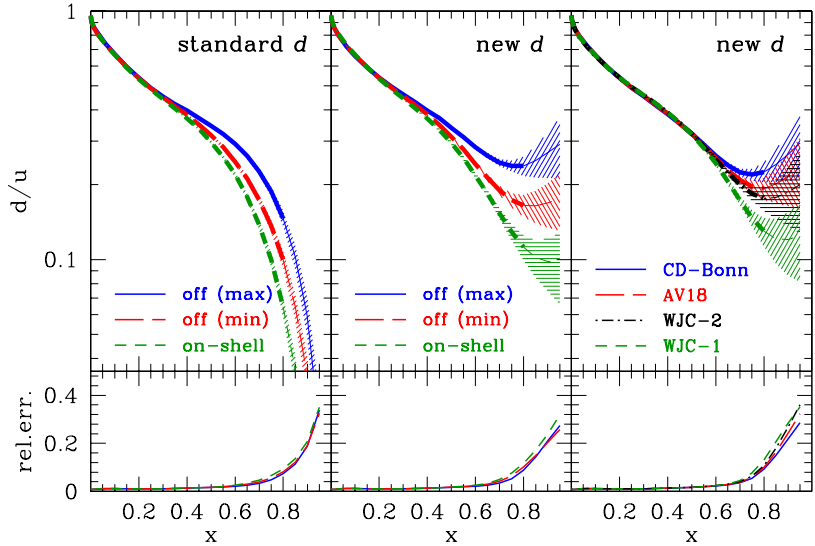}
}
\caption{The CJ11 $d/u$ ratio at $Q^2=10$ GeV$^2$ obtained with different
 	$d$ quark parametrization and nuclear corrections:
   ({\it left}) standard $d$ quark \protect\cite{Accardi:2009br}
 	and AV18 deuteron wave function for several off-shell
 	corrections models;
   ({\it middle}) modified $d$ quark parametrization;
   ({\it right}) dependence on the wave function
 	for a fixed off-shell correction (mKP).
 	The shaded bands and bottom panels show the $\Delta\chi=1$
 	(tolerance $T=1$) absolute and relative PDF errors, respectively.  
      For details, see the original paper \protect\cite{Accardi:2011fa}.}
\label{fig:d2uparams}
\vskip-.1cm
\end{figure}

A further source of theoretical uncertainty was investigated by utilizing a more flexible parametrization for the valence $d_v$ quark at large-$x$, which now includes a small admixture of the valence $u_v$ PDF,
\begin{align}
  d_v(x) \rightarrow d'_v(x) = 
    a_0^{d} \Big[ d_v(x) / a_0^{d} + b\, x^c u_v(x) \Big],
\label{eq:newd}
\end{align}
where $a_0^{d}$ is the $d$ quark normalization, and $b$ and $c$ are
two additional parameters\cite{Accardi:2011fa,Owens:2012bv}. The result is that the $d/u$ ratio at $x \to 1$ can now span the range $[0,\infty)$ rather than being limited to either 0 or $\infty$ as in all previous PDF fits.
A finite, nonzero value of this ratio is in fact expected from several nonperturbative models of nucleon structure\cite{Jimenez-Delgado:2013sma,Holt:2010vj}. 
It is also required from a purely practical point of view because it avoids potentially large biases on the fitted $d$ quark PDF, as illustrated in the left and middle panels of Figure~\ref{fig:d2uparams} and discussed in more detail in Section~\ref{sec:theorybiases}.

\section{Applications}

Let us now discuss in turn a hadronic physics and a collider physics application of the fitted CJ12 PDFs, and explore the consequences of the nuclear uncertainties in either case.

The ratios of the $d$ to $u$ PDFs for the three CJ12 sets are shown in Figure~\ref{fig:CJ12d} right. These are constrained up to $x\approx 0.8$ by the enlarged data set considered in the CJ fits, but can be confidently extrapolated to $x=1$ thanks to the modified $d$ quark parametrization \eqref{eq:newd}.
As the magnitude of the nuclear corrections increases, the $d/u$ intercept at $x=1$ rises\cite{Owens:2012bv}. Including the PDF errors, it was found that 
\begin{align}
  d/u \,\xrightarrow[\,x\rightarrow 1\,]{} \, 0.22
	\pm 0.20 \, \text{({\small PDF})}
	\pm 0.10 \,\text{(nucl)},
\label{eq:dulimit}
\end{align}
where the central value is obtained as an average of the CJ12max and CJ12min PDFs, the first error is from the PDF fits, and the second error reflecting
the theoretical nuclear uncertainty is obtained by considering the difference between the CJ12min and CJ12max PDFs.  These values encompass the full 0-0.5 range of available theoretical predictions\cite{Feynman:1973xc,Close:1973xw,Farrar:1975yb,Melnitchouk:1995fc,Wilson:2011aa}.  
However, it is also clear that a relatively modest
improvement in statistical precision and reduction of nuclear
uncertainty would allow one to restrict the range of allowable
physical mechanisms\cite{Roberts:2013mja}.

\begin{figure}[t]
\center
\includegraphics[width=0.95\linewidth,bb=18 485 565 700,clip=true]
                {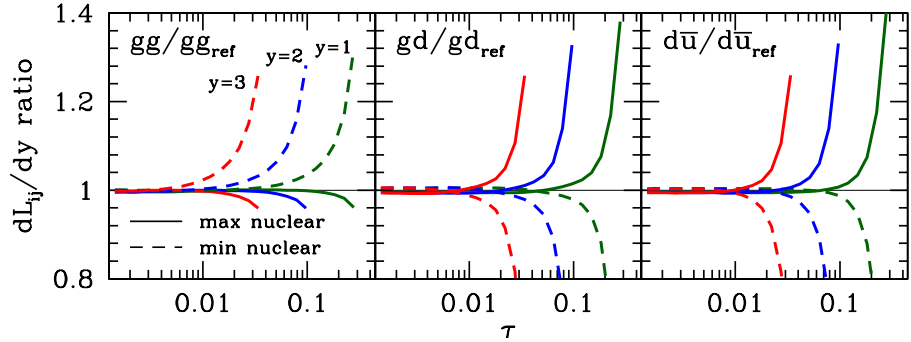}
\vskip-.3cm
\caption{CJ11  differential parton luminosities for
	$gg$, $gd$ and $d\bar u$ parton collisions
	at fixed rapidity $y=1,2$ and 3, as a function of
	$\tau = \sqrt{\hat s / s}$, illustrating the variations
	due to the choice of nuclear corrections 
        \protect\cite{Accardi:2011fa}.}
\label{fig:lum_y}
\end{figure}

\begin{figure}[t]
\parbox{0.6\linewidth}{
\rotatebox{-90}{\includegraphics[height=0.94\linewidth,clip=true]{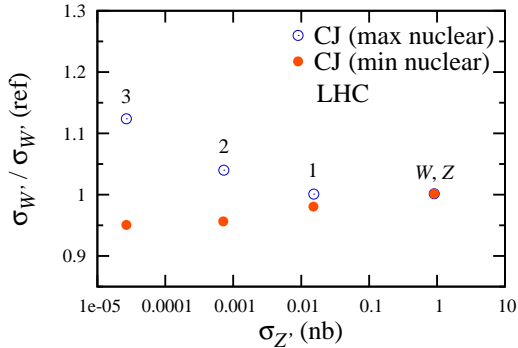}}
}
\parbox{0.39\linewidth}{
\caption{Ratio of $W'$ boson 
	$\sigma_{W'} = \sigma_{W'^+} + \sigma_{W'^-}$ cross section,
	versus the $Z'$ boson cross section $\sigma_{Z'}$ for varying
	$Z'$ boson masses (in TeV next to
	the points). The cross sections are computed for $\sqrt s = 7$ GeV 
        from CJ11 PDFs  \protect\cite{Accardi:2011fa} with minimum
	(filled red circle) and maximum (open blue circle) nuclear
	corrections, relative to the reference cross section calculated
	from the central CJ11
        PDFs\protect\cite{Brady:2011hb}.\newline\newline }
\label{fig:WZprime_incl}
}
\vskip-0.5cm
\end{figure}

The large nuclear uncertainty in the quark PDF (and that in the
gluon PDF arising from jet-data induced anticorrelation with the $d$ quark\cite{Accardi:2011fa}), have also potentially profound implications for collider experiments. To illustrate this in general terms, consider the differential parton luminosities for production of an object of mass $\sqrt{\hat s}$ at rapidity $y$ in a hadronic collision of center of mass energy $\sqrt s$, 
\begin{align}
dL_{ij} / dy
= \big[ f_i\left( x_1, \hat{s} \right) 
	f_j\left( x_2, \hat{s} \right) 
        + (i \leftrightarrow j)
  \big] / \big[ s (1+\delta_{ij}) \big] ,
\end{align}
where $x_{1,2} = \tau e^{\pm y}$ with $\tau = \sqrt{\hat s / s}$. 
The differential luminosities, normalized to the reference fit calculation, are shown in Figure~\ref{fig:lum_y} as a function of 
$\tau$ for three values of rapidity. (Since the ratios are largely 
independent of the hard scattering scale, these plots are also independent of $s$). 
The sensitivity to large $x$ PDFs, hence to nuclear uncertainties, grows the larger the mass of the produced object and the larger its rapidity at a given mass. For example, the nuclear uncertainty becomes relevant for $W$ production at rapidity larger than 2  at the Tevatron, and larger than 3.5 at the LHC\cite{Brady:2011hb}, as we shall see in more detail.
For particles of heavier mass, such as the putative $W'$ and $Z'$ bosons, the nuclear uncertainty at large $x$ may become large also in the inclusive production cross section\cite{Brady:2011hb}, as illustrated in Figure~\ref{fig:WZprime_incl}.  
As the current LHC limits put the $W'$ and $Z'$ masses approximately above 2.5 TeV for Standard-Model like couplings\cite{Dissertori:2012cra}, one can appreciate how nuclear uncertainties and other large-$x$ theoretical uncertainties may significantly affect the interpretation of signals of new particles and an accurate measurement of their properties.

\section{Theoretical biases at $x \rightarrow 1$}
\label{sec:theorybiases}

\begin{figure}
\center
  \vskip-.45cm
  \includegraphics[width=0.89\linewidth,trim=20 250 20 0,clip=true]
                  {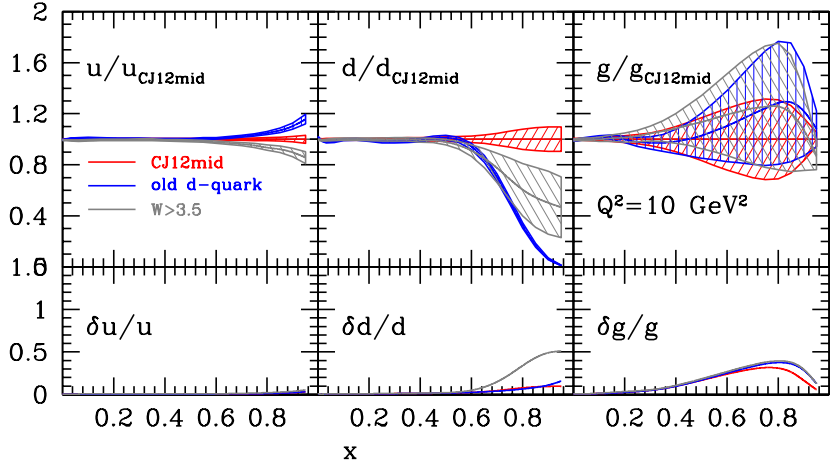}
  \vskip-.3cm 
  \caption{Comparison of PDFs obtained with a conventional $W>3.5$~GeV
  cut, and using the old $d_v$ quark parametrization with $W>1.7$~GeV
  to the CJ12mid set. A tolerance $T=1$ is used for clarity.}

\label{fig:CJ12-udg}
\end{figure}

\begin{figure}[t]
\center
  \vskip-.2cm
  \includegraphics[width=0.43\linewidth,trim=0 120 120 0,clip=true]
                  {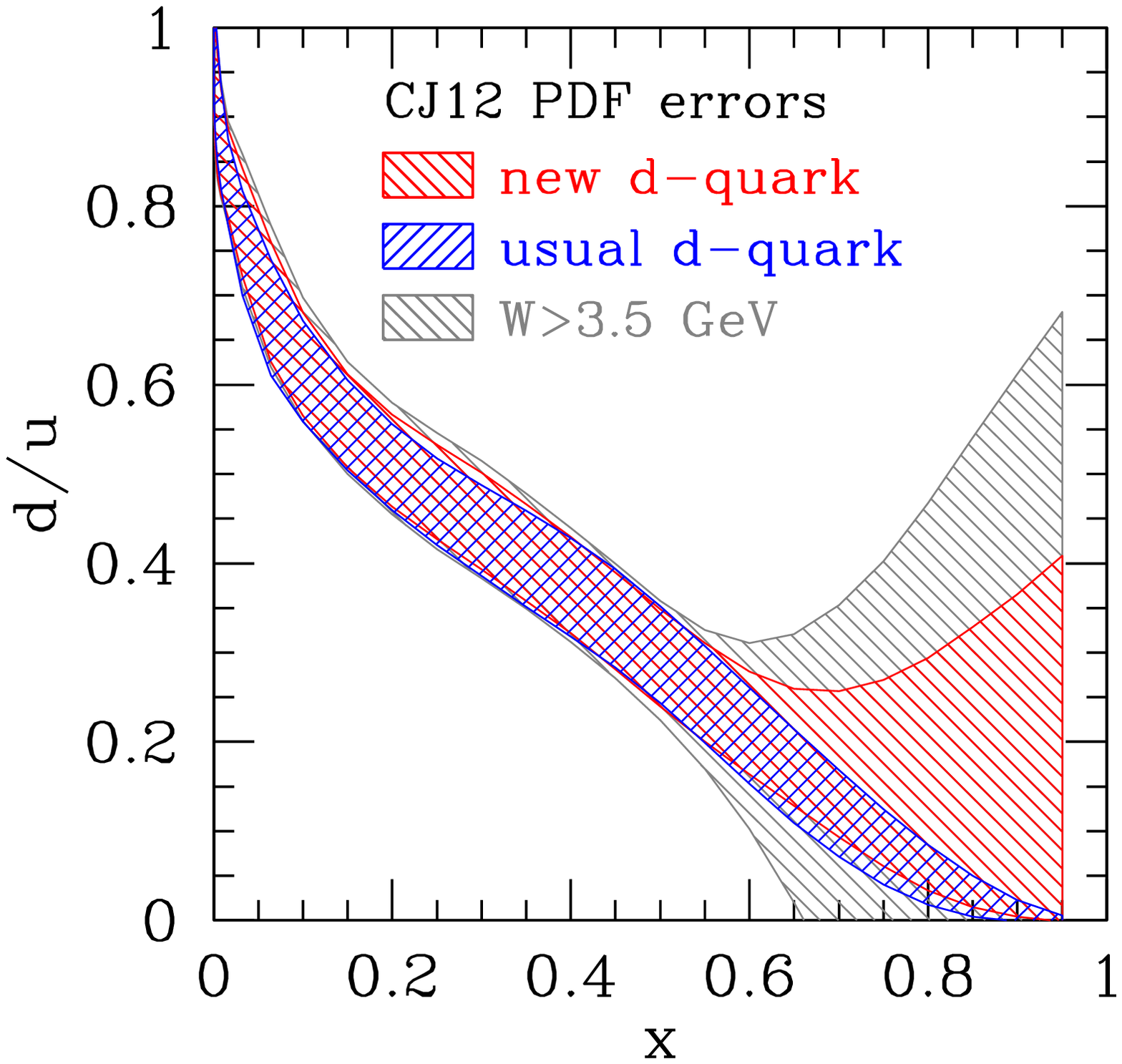}
  \hspace{.5cm}
  \includegraphics[width=0.43\linewidth,trim=0 120 120 0,clip=true]
                  {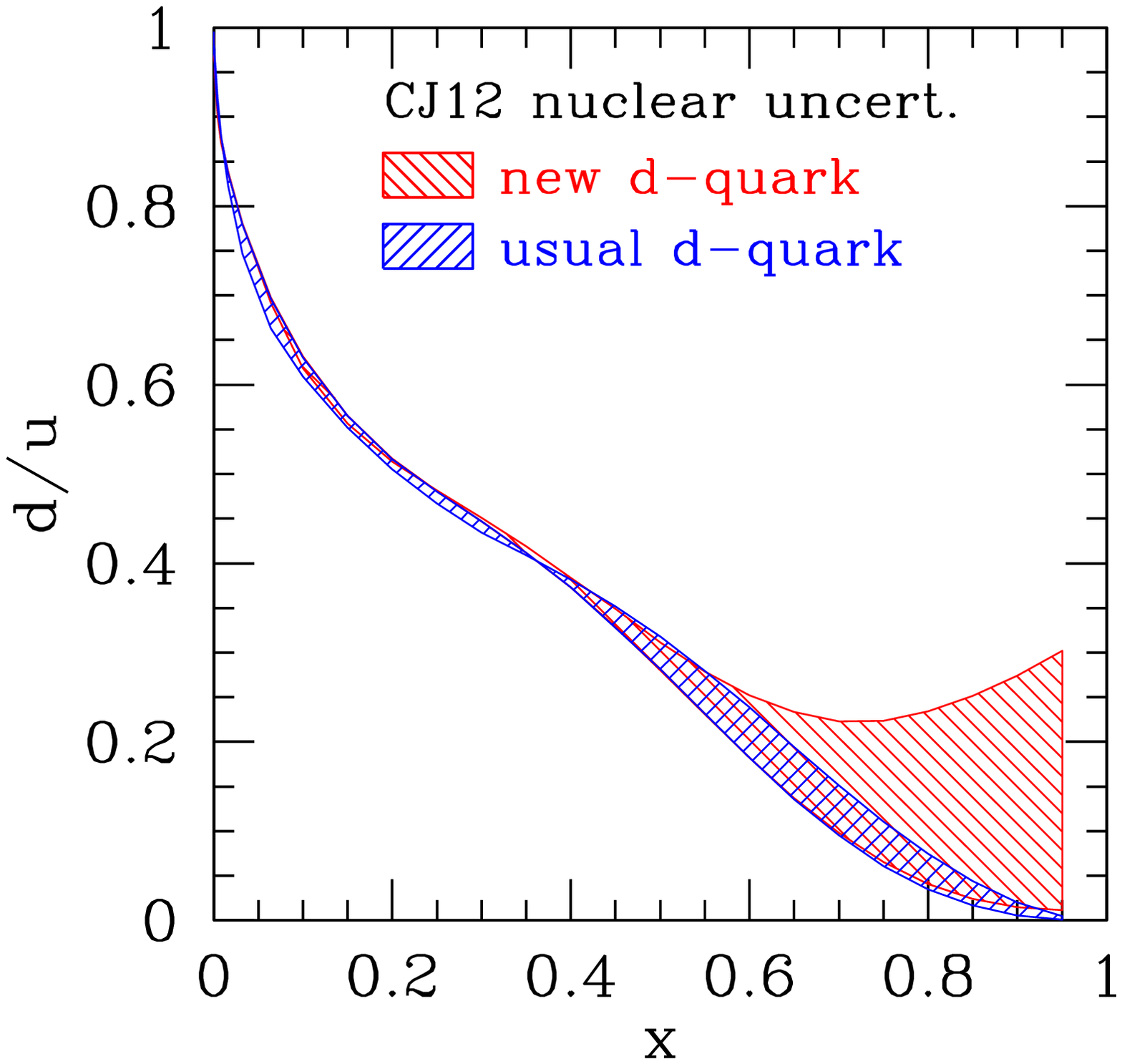}
  \vskip-.2cm                
  \caption{
    PDF errors with tolerance $T=10$ and nuclear uncertainties
    for the fits in Fig.~2.
   }
\label{fig:d-biases}
\end{figure}

The importance of using a more flexible $d$-quark parametrization than
in traditional fits cannot be overemphasized. Figures~\ref{fig:CJ12-udg}
and \ref{fig:d-biases} show a comparison of the CJ12mid fits to a fit
obtained with the standard $W>3.5$~GeV cut, and a fit obtained with
$W>1.7$~GeV but standard $d$-quark parametrization (namely, $b=c=0$
in Eq.~\eqref{eq:newd}).

In Figure~\ref{fig:CJ12-udg}, the ratios of the $u$, $d$ and gluon
PDFs in the two new fits to the CJ12mid fit are presented with a
tolerance $T=1$ (for clarity).  The fit using the standard
parametrization starts deviating from the CJ12mid fit at
$x \gtrsim 0.5$.  This behavior is clearly not data driven
(it starts already inside the covered kinematic range), but is
forced by the use of the functional form $d \propto (1-x)^{a_d}$
rather than the more flexible parametrization \eqref{eq:newd}.
The dramatic decrease of the standard $d$ quark PDF towards 0 can
be compensated in the fit by a slightly increased $u$ quark and gluon distributions, due to the correlation induced by the large-$x$ DIS data and jet data, respectively.

Further insight can be obtained by considering the $d/u$ ratio in
Figure~\ref{fig:d-biases}.  In this case the standard parametrization
forces $d/u$ to take either the value of 0 or $\infty$.
In contrast, the extended $d'$ parametrization allows for limiting
values in the entire range between 0 and $\infty$.  In the CJ12 fit,
the data does not seem to warrant the behavior of the standard 
parametrization, which lies at the edge of the PDF error band
of the CJ12mid fit. Not only does the standard $d$
parametrization underestimate the central fitted value at
$x \gtrsim 0.5$, it also underestimates the PDF uncertainty,
as well as the nuclear uncertainty. As a consequence one could risk, {\it e.g.},  to interpret as signal of new physics what instead is an artificially small calculated cross section, as it has already happened in the past in analogous situations\cite{Huston:1995tw,Kuhlmann:1997wd}.

Other potential biases exist at large $x$, related to theoretical
corrections not yet included included in the perturbative QCD
calculations utilized in the CJ12 analysis.  These include large-$x$
resummation\cite{Anderle:2012rq,Courtoy:2013qca}, jet mass corrections\cite{Accardi:2008ne}, and higher-order terms in the perturbative expansion\cite{Blumlein:2008kz}.  However, these typically scale as $1/Q^2$ (or resemble such a power correction at low $Q^2$) and will mainly affect the extraction of the HT term, leaving the
leading-twist PDFs largely unchanged, as was found, {\em e.g.}, for the model
dependence of target mass corrections\cite{Accardi:2009br}.
This makes the choice of parametrization possibly the largest remaining
theoretical bias in the determination of the $d$-quark PDF at
large $x$.  A full theoretical unbiasing should be pursued by
generalizing the $d'$ quark functional form adopted in the CJ12
fits and investigating the related quantitative extraction of
$d$-quark PDF errors at $x \to 1$.

\section{Constraining the nuclear corrections}

Given the current size of nuclear uncertainties and their impact on phenomenology across low-energy hadron physics and high-energy particle physics, it is imperative to reduce this source of theoretical uncertainty.

On the theory side the main difficulty resides in the interplay of the hard interaction at the nucleon level, described in terms of partonic degrees of freedom and calculable directly from the QCD Lagrangian, and the comparatively soft nucleon-nucleon interactions described in terms of hadronic degrees of freedom with effective interactions fitted to nucleon elastic scattering data. Oftentimes the soft nuclear and hard nucleonic interactions are assumed to factorize, even though this is not formally guaranteed the way QCD factorization theorems allow to separate the hard partonic interactions from nonperturbative PDFs\cite{Jaffe:1985ci}.
Moreover, the medium modifications of the nucleons are almost unknown experimentally, especially in the nucleon momentum range relevant for inclusive DIS. Therefore, theoretical models of nuclear corrections seem at present irreducibly unconstrained, and comparison to non-inclusive DIS experimental data is mandatory for further theoretical progress.

If, however, one's purpose was solely to obtain a precise enough but nuclear uncertainty free extraction of the $d$ quark, one would need new experimental data to compensate for the loss in statistics deriving from excluding the deuteron data shown in Figure~\ref{fig:CJ12d} left. This new data can come from 2 sources. The first is data on nuclear targets, but with observables designed to minimize the impact of nuclear corrections, for example, DIS on deuteron targets with detection of a slow spectator proton, which guarantees that the electron scatters on a quasi-free neutron, or inclusive DIS on 3-Helium and Tritium, where the nuclear effects largely cancel in the ratio of the respective cross sections (future BONUS 12 and MARATHON experiments at Jefferson Lab, respectively\cite{MARATHON,BONUS12}). The second source is weak interactions in proton target collisions.
These contribute, through photon-$Z$ boson interference, to electron and positron DIS cross sections at large $Q^2$ only (for which, unfortunately, the combined run-I HERA data\cite{Aaron:2009aa} have insufficient large $x$ coverage), but are of leading order in parity violating charge asymmetries (recent limited large-$x$ coverage $F_{2,3}^{\gamma Z}$ measurements at HERA\cite{Aaron:2012qi,Abramowicz:2012bx}, and future SOLID\cite{SOLID} and PVDIS\cite{PVDIS} experiments at Jefferson Lab). In principle, weak charged currents can also be directly measured in neutrino and anti-neutrino scattering on proton targets, for which however the old WA21/25 data\cite{Jones:1987gk,Jones:1994pw} cannot be easily used in global fits, and no new data is planned for the foreseeable future. 
This leaves one with $W$ and $Z$ boson production in $p+\bar p$ collisions at Tevatron and $p+p$ collisions at RHIC and the LHC, which I shall focus on next.

\begin{figure}[t]
\center
\rotatebox{-90}{\includegraphics[height=0.5\linewidth]{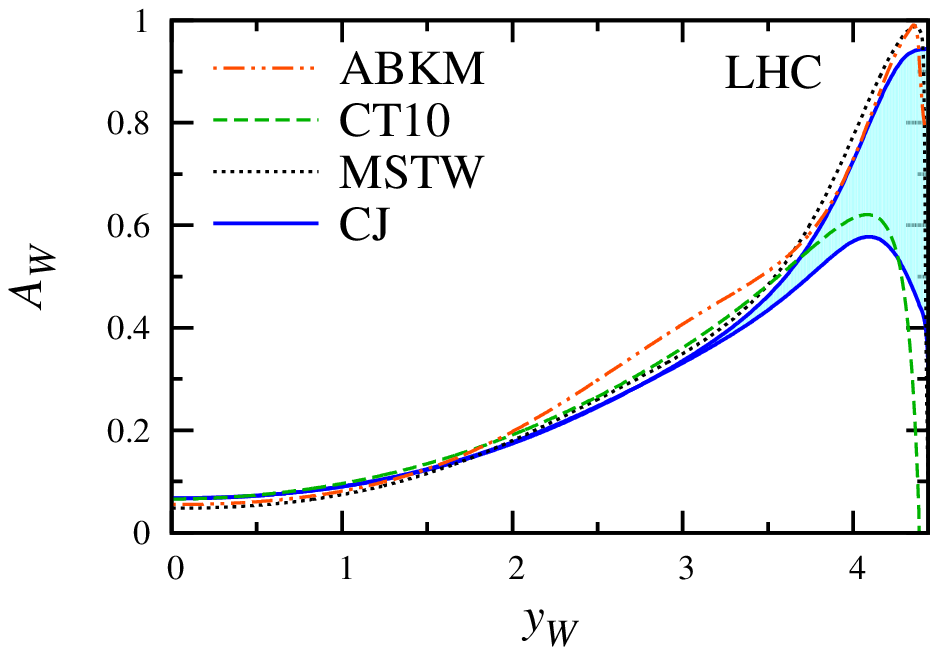}}
\hspace*{-0.5cm}
\rotatebox{-90}{\includegraphics[height=0.5\linewidth]{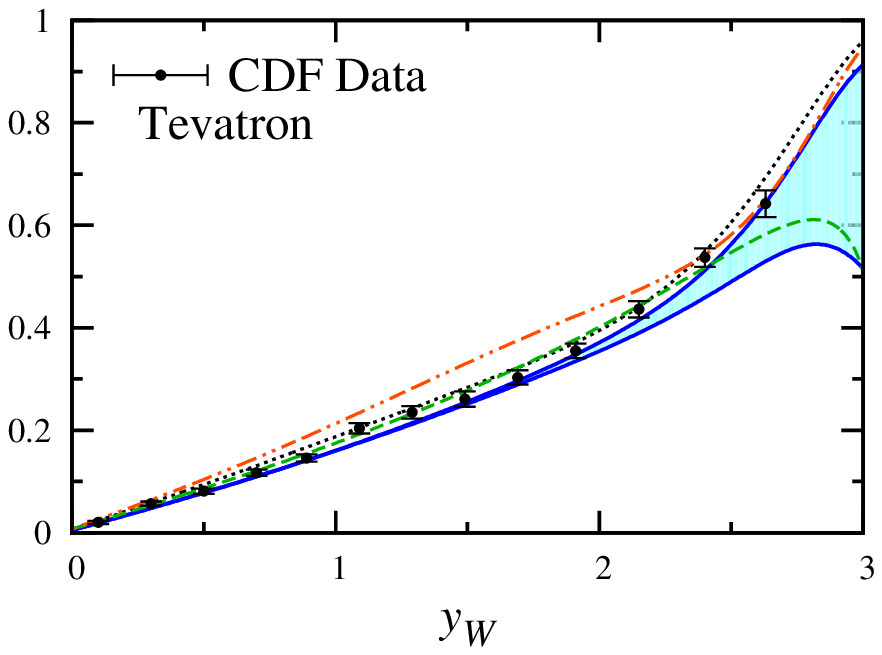}}
\vskip-.3cm
\caption{$W$ boson asymmetry $A_W$ as a function of the $W$ rapidity
	$y_W$ at LHC and Tevatron,
	computed from CJ11 PDFs with minimum (upper blue solid)
	and maximum (lower blue solid) nuclear 
        corrections~\protect\cite{Brady:2011hb}.
	The asymmetries using the ABKM, CT10 and MSTW
	PDF sets are also shown.}
\label{fig:W_asym}
\end{figure}

As discussed at the end of Section~\ref{sec:overview}, global QCD fits allow for the intriguing possibility to combine deuteron and proton target data to obtain experimental constraints on the nuclear models, and at the same time fully utilize the available statistics of the nuclear target data. This is very well demonstrated by considering the directly reconstructed $W$ asymmetry at Tevatron. As shown in Figure~\ref{fig:W_asym}, this process is very sensitive to nuclear corrections at $y\gtrsim 2$, and the comparison\cite{Brady:2011hb} with the very precise CDF data\cite{Aaltonen:2009ta}, as well as the total $\chi^2$ values in the CJ12 fits, suggests that nuclear effects are somewhere between the minimum and central ones considered in the CJ12min and CJ12mid fits, respectively. This was also confirmed in a similar analysis by Martin {\it et al.}\cite{Martin:2012da}, where however the nuclear correction was assumed to be $Q^2$ independent and directly fitted to data rather than calculated in a model. On the phenomenological side, this result disfavors nonperturbative proton models based on the SU(6) spin-flavor symmetry, which predict $d/u \rightarrow 1/2$ as $x\rightarrow 1$. More importantly, it exemplifies the power of global fits in combining data from across the board, and is the first step in establishing the experimental foundation that was until now missing for a qualitative jump in our ability to theoretically understand and describe high-energy processes in nuclei -- and this with proton targets! Interestingly, this procedure is not limited to deuteron targets, but can in principle be adapted to provide a new line of experimental constraints of nuclear effects in heavier nuclei, thus shedding new light on the EMC effect that for more than 30 years has eluded a satisfying theoretical explanation\cite{Higinbotham:2013hta,Norton:2003cb,Frankfurt:2012qs}.

The $W$ charge asymmetry at CDF unfortunately seems the only observable that currently has this potential. The asymmetry in lepton production from $W$ decays has insufficient large $x$ coverage due to decay vertex smearing, while $Z$ rapidity distributions from CDF and D\O\ have insufficient precision\cite{Owens:2012bv}. At the LHC one would need measurements with better than 10\% precision at $y\gtrsim 3.5$, namely, at the edge of the LHCb acceptance for $\sqrt{s}=7$ TeV, and the experimental accuracy of the $W$ asymmetry is inherently reduced due to the difference between $p+p$ scattering compared to $p+\bar p$ \cite{Brady:2011hb,Lohwasser:2010sp}. Nonetheless, inclusion of the available LHC data in the CJ fits is underway to quantify its impact on nuclear uncertainty constraints. 

The potential of large rapidity $W$ and $Z$ measurements in proton-proton collisions at the RHIC\cite{Bourrely:1993dd,Melnitchouk:1996fh,Surrow-DIS2013,Bourrely:2013qfa} should also be explored. 
In particular\cite{W_at_RHIC_private}, the STAR detector can in principle provide directly reconstructed $W$ asymmetries at pseudorapidity $-1\leq \eta \leq 1$ using the central detector, accessing the $d$ quarks at $0.05<x<0.5$, and possibly up to $\eta \approx 2$ using the end cap, which will access the $x>0.5$ region we have been focusing on in this review. Although this is a challenging measurement\cite{Lohwasser:2010sp}, it would provide one with an important validation of the CDF measurement\cite{Aaltonen:2009ta}, as well as a test of its energy dependence. 
Of further interest would be measurements of $W$, $Z$ and dilepton production in proton-deuteron collisions at large rapidity in the deuteron direction\cite{Kamano:2012yx,Ehlers_etal}: these would access large-$x$ quarks in the deuteron and test theoretical nuclear corrections in a complementary way to electron-deuteron DIS measurements.

Looking at the future, one could also envisage utilizing the sensitivity of large-$x$ gluons to nuclear corrections (induced by their correlation with  $d$-quark in jet data) to further constrain nuclear models analogously to what I am proposing here for the $d$ quark. This would require a further observable sensitive to large-$x$ gluons in protons, {\it e.g.}, data on top quark production from Tevatron and LHC, which is directly sensitive to large-$x$ gluons\cite{Czakon:2013tha}, and furthermore free from the subleading in $1/Q^2$ corrections that limit the sensitivity of longitudinal $F_L$ structure function or cross section measurements in fixed target DIS experiments. 
The planned Electron-Ion Collider\cite{Boer:2011fh,Accardi:2012hwp} and Large Hadron-electron Collider\cite{AbelleiraFernandez:2012ty} projects will be able to perform a complete set of structure function measurements for gluon and quark flavor separation at large $x$ (and large $Q^2$).
Finally, new experiments focused on large rapidity particle production, such as the proposed AFTER@LHC experiment\cite{Brodsky:2012vg}, would provide invaluable data to not only obtain precise large-$x$ PDFs, but also and more in general to investigate QCD in the nuclear medium.

\section{Conclusion}

The recent CTEQ-Jefferson Lab collaboration investigations, culminating in the public CJ12 PDF release\cite{Owens:2012bv,CJweb,CJ-LHAPDF}, have demonstrated the intimate interconnection of hadronic and high-energy physics, exemplifying one of the across-the-board connections called for in the opening quote by C.~Quigg for further progress in these fields. Namely, global QCD fits have become capable of constraining theoretical models of nuclear corrections in the deuteron (as well as in heavier nuclei).  Not only will this reduce the nuclear uncertainty on the fitted PDFs with important phenomenological consequences on physics ranging from nonperturbative proton structure to beyond the standard model interactions, it will also provide a new avenue for progress in the theoretical understanding of high-energy processes involving \emph{nuclei}, using weak interactions on \emph{proton} targets from Jefferson Lab to the LHC.

\vskip.6cm\noindent
{\bf Acknowledgments:} 
I am very grateful to my CJ colleagues for the enjoyable and productive collaboration. While the opinions expressed in this paper are the author's responsibility only, interesting discussions with E.Aschenauer, S.Forte, J.Rojo, E.Sichtermann, and R.Thorne are gratefully acknowledged. 
This work was supported by the DOE contract No.~DE-AC05-06OR23177,
under which Jefferson Science Associates, LLC operates Jefferson Lab, and by the DOE contract No. DE-SC008791.

\end{document}